\documentclass[aps,prb,superscriptaddress,showpacs,showkeys]{revtex4}

\usepackage{amsmath}
\usepackage{amsfonts}
\usepackage{bm}
\usepackage{graphicx}
\usepackage{color}
\usepackage{hyperref}

\newcommand{\vk}{\bm k}

\newcommand {\mv}[1]{\langle #1 \rangle}
\newcommand {\vrv}{\mbox{\boldmath$r$}}

\begin{document}

\title{Admittance of planar two-terminal quantum systems}

\author{U. Wulf}
\affiliation{Technische Universit\"at Cottbus, Fakult\"at 1, Postfach 101344,
             03013 Cottbus, Germany }
\affiliation{IHP/BTU JointLab, Postfach 10 13 44, 03013 Cottbus, Germany}
\email{wulf@physik.tu-cottbus.de}

\author{P. N. Racec}
\affiliation{IHP/BTU JointLab, Postfach 10 13 44, 03013 Cottbus, Germany}
\affiliation{National Institute of Materials Physics, PO Box MG-7,
76900 Bucharest Magurele, Romania}
\email{racec@physik.tu-cottbus.de}

\author{E. R. Racec}
\affiliation{Technische Universit\"at Cottbus, Fakult\"at 1, Postfach 101344,
             03013 Cottbus, Germany }
\affiliation{University of Bucharest, Faculty of Physics, PO Box MG-11,
76900 Bucharest Magurele, Romania}
\email{roxana@physik.tu-cottbus.de}

%


\begin{abstract}
We develop an approach to calculate the admittance of
effectively one-dimensional open quantum systems
in random phase approximation. 
The stationary, unperturbed system is 
described within the Landauer-B\"uttiker formalism
taking into account the Coulomb interaction  in the Hartree approximation.
The dynamic changes in the effective potential are calculated
microscopically from the charge-charge correlation function resulting from the
stationary scattering states.
We provide explicit RPA-expressions
for the quantum admittance. 
As a first example the case of a 
quantum capacitor is considered where we can derive
a small-frequency expansion for the admittance
which lends itself to an experimental testing of the theory.
A comparison of the low-frequency expansion 
with the complete RPA-expression shows that for a 
quantum capacitor a simple classical equivalent circuit with 
frequency-independent elements does not
describe satisfactorily the quantum-admittance with increasing the frequency.
\end{abstract}

\maketitle

\section{Introduction}

The ac-transport properties of nearly ballistic devices
are interesting from the point of view of basic research
as well as from  the point of view of technological 
applications. In basic research ac-transport
can provide valuable
extra information to stationary transport.
In applications impedances are of crucial importance
for the lay-out of ac-circuits.

In the past years a number of methods for the 
description of time-dependent transport phenomena in mesoscopic systems
have been proposed (for a recent review see Ref. \ [\onlinecite{platero}]).
Among these approaches are techniques
based on non-equilibrium Greens functions,\cite{lopez,you,Zhou,faleev}
Wigner functions,\cite{aronov98} bosonization schemes,\cite{safi} 
and phenomenological considerations.\cite{feiginov01}
If there are only small ac-fields
it seems promising to calculate the linear ac-response
to a perturbation of the stationary system
described in the successful Landauer-B\"uttiker formalism (LBF).
\cite{lanbue1,lanbue2,lanbue3,lanbue4,lanbue5,lanbue6}
Such an approach  has been developed
in Refs. \ [\onlinecite{buettiker93a,buettiker93b,
buettiker93c,buettiker94a,buettiker96a,buettiker96b}] and it 
has been used in a number of applications.
\cite{christen96a,christen96b,pretre96,gopar96,brouwer97,
blanter_pr00,hekking}
Here the response to an external potential is derived 
'which prescribes the potentials $(U_\alpha)$ in the reservoirs only' (see 
Ref.\ [\onlinecite{buettiker93a}]). 
The reservoirs carry the charge $Q_\alpha$ so that the
ac-potential leads to a perturbation in the Hamiltonian as given by
$H_1 = \sum_\alpha U_\alpha Q_\alpha$. 
This approach avoids the
calculation of the time-dependent
potential within the scattering area, i. e. outside the
contact reservoirs. However, the knowledge of the microscopic potential
in the scattering area is necessary to derive a formal response theory,
beyond the  often invoked spatially uniform electric field 
perturbation.\cite{fu}
As pointed out in Refs. \ [\onlinecite{blanter98a,blanter98b,hekking}]
the appropriate response formalism for the interacting electron system is the 
random phase approximation (RPA). In our previous papers 
Refs. \ [\onlinecite{emrs03,emrs05}] we demonstrated
the application of the complete RPA-scheme 
to open  stationary systems described in the Landauer-B\"uttiker formalism.
A complete RPA-scheme\cite{kadanoff} requires as a central element
the calculation of the irreducible  polarization
$\Pi_0(\bm{r}, \bm{r}', \omega)$ from the self consistent
 scattering functions of the stationary system. 
As a second necessary ingredient for the implementation of the complete 
RPA-scheme, it was shown in  Refs. \ [\onlinecite{emrs03,emrs05}]
how the dynamic total potential in 
the scattering area can be determined
microscopically using
the calculated irreducible  polarization and the Greens function for the
Poisson equation with Dirichlet boundary conditions.

After a formal derivation of our  theoretical
approach we   derive in this paper 
explicit RPA-expressions  for the frequency dependent impedance
in a general two-terminal device under large dc bias. 
These expressions are evaluated for the case of a quantum capacitor.
In the limit $\omega \rightarrow 0$ an expansion of the admittance
 follows as given by
\begin{equation}
Y =   -  i \omega (Y_1 +   i \omega Y_2 +\dots),
\label{lowfri}
\end{equation}
with real constants $Y_1$ and $Y_2$ that can be calculated directly
from the scattering functions of the stationary system.
In numerical computations we determine the admittance 
of a  MIS-type heterostructure\cite{emrs05} 
on which measurements of the static capacitance have already been 
made.\cite{dolgopolov}
In order to propose an experimental test of our RPA-approach
we first compute the drain-source-voltage dependence of the coefficients
$Y_1$ and $Y_2$ in the limit  $\omega \rightarrow 0$.
For higher frequencies we find numerically, first,
 that the expansion in Eq.\ (\ref{lowfri})
becomes invalid very quickly and, second,
that an equivalent circuit
with frequency independent R and C elements does not reflect correctly
the ac-properties of the considered system. Instead,
there are pronounced and systematic
deviations from the equivalent circuit behavior
which  should be testable in experiments as well.

\section{General admittance formula}

\subsection{Stationary system}
\label{stat_LB}

\begin{figure}[ht]
\noindent\includegraphics[width=3.in]{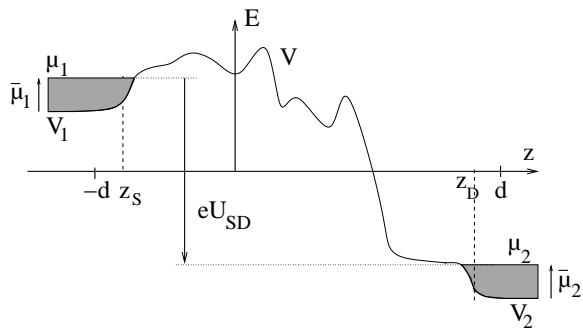}
\caption
{Sketch of the potential energy along the growth direction in a 
planar two-terminal system. It can be seen
that $-eU_{SD} = V_2 + \bar{\mu}_2 - V_1 -  \bar{\mu}_1$,
where $\bar{\mu}_2$ and $\bar{\mu}_1$ denote the difference
between the chemical potential and the bottom of the
conduction band in the bulk contact. }
\label{fig1}
\end{figure}
We consider a planar two-terminal system with
a source contact ($s=1$, $z<z_S$) and a drain contact ($s=2$, $z>z_D$) 
under an external bias $U_{SD}$
as shown schematically in Fig.\ \ref{fig1}. 
The material of the contacts can either be
a strongly $n$-doped semiconductor or a metalization.
In the mean field theory the effective potential energy is independent of the
perpendicular coordinate ${\vrv}_{\perp} = (x,y)$,
$V=V(z)$. 
Because of the effective screening the potential energy is constant in the
bulk of the source contact, $V(z<-d) = V_1$, and in the bulk
of the drain contact,
$V(z>d) = V_2$. 
We write for the wave function\cite{roxana}
\begin{equation}
\varphi({\bm r})  = 
 \psi^{(s)}(\epsilon,z)
\frac{\exp{ (i{\bm k}_{\perp} {\bm r}_{\perp}) }}{ \sqrt{A} },
\label{6ansatz}
\end{equation}
where ${\vk}_{\perp} = (n_x 2 \pi /L_x, n_y 2 \pi /L_y)$, 
$n_x$ and $n_y$ are integer numbers,  
and $A=L_xL_y$ is the cell area
of periodic boundary conditions in the perpendicular directions.
It follows from the time-independent Schr\"odinger equation
$(\hat{H}_0-E) \varphi =0$, with $\hat{H}_0=-\hbar^2/(2m^*)\Delta + V(z)$, 
that 
\begin{equation}
\left[- \frac{\hbar^2 }{2m^* } \frac{d^2}{dz^2} + V(z) - \epsilon
\right]  \psi^{(s)}(\epsilon,z)  = 0,
\label{6schroeff}
\end{equation}
with 
\begin{equation}
E = \epsilon + \frac{\hbar^2}{2m^*}k_{\perp}^2 .
\label{totalE}
\end{equation}
The wave function in Eq.\ (\ref{6ansatz}) is defined 
by the set of quantum numbers
$(s,\epsilon,{\bm k}_{\perp})$ 
without considering the spin quantum number.
Consistent with  the Landauer-B\"uttiker formalism
the $\psi^{(s)} (\epsilon,z)$ are scattering functions,
where the index $s$ defines the direction of incidence:
the scattering functions incident from the source
contact ($s=1$) exhibit the asymptotics 
\begin{equation}
\psi^{(1)}(\epsilon,z) =\frac{\theta(\epsilon-V_1)}{\sqrt{2\pi}}
\left\{  
\begin{array}{ll}
\exp{[i k_1 (z+d)]} +{\bm S}_{11}(\epsilon) \exp{[-i k_1 (z+d)]},
& \mbox{~~for~~} z \le -d \\
& \\
{\bm S}_{21}(\epsilon) \exp{[i k_2 (z-d)]},
& \mbox{~~for~~} z \ge d,
\end{array}
 \right. 
\label{6psi-out-1}
\end{equation}
where ${\bm S}_{11}$ and ${\bm S}_{21}$ are elements of the energy dependent 
$2\times 2$ scattering matrix,\cite{roxana}
\begin{equation}
k_s(\epsilon) = \sqrt{\frac{2 m^*}{\hbar^2} (\epsilon - V_s)}, 
\label{6k}
\end{equation}
and $\theta(\epsilon-V_1)$ is the step function.
Analogous expressions hold for the drain-incident scattering functions 
$\psi^{(2)}(\epsilon,z)$. 

As usual, the effective (total) potential energy contains
an external part,
$V_{ext}$, and a second part, $V_{el}$, coming from the Coulomb
interaction between the electrons. 
The external potential arises typically
from different band offsets in the used materials
or from fixed external charges like fully ionized impurities. 
The Coulomb interaction between the electrons is taken in the Hartree 
approximation
so that $V_{el}$ obeys the Poisson equation
having as sources only the electronic charge density
\begin{equation}
\frac{d^2}{dz^2} V_{el}(z) = -\frac {e^2}{\kappa_s} \rho(z),
\label{poisson}
\end{equation}
where $\kappa_s$ is the dielectric constant of the host material. 
In Appendix \ref{appendix6} we  reproduce that the electron density
\begin{equation}
\rho(z) =
       2 \frac{m^*}{2 \pi \beta \hbar^2}
         \sum_{s=1,2}  \int_{V_s}^{\infty} d \epsilon \; g_s(\epsilon)
         \left| \psi^{(s)}(\epsilon,z) \right|^2
         \ln { \left\{ 1+ \exp {\biglb( \beta (\mu_s - \epsilon) \bigrb)} \right\} }
                            \label{6qq-2}
\end{equation}
in a formal
quantum statistical approach needed 
to formulate the linear response theory which we will describe in the next 
section.
In Eq.\ (\ref{6qq-2}) $g_s(\epsilon)= m^*/ [\hbar^2 k_s(\epsilon)]$
is the one-dimensional density of states,
$\beta =1/(k_BT)$, and
we included a factor of two to account for the spin degeneracy. 

\subsection{Harmonic perturbation}
\label{harmpert}

\subsubsection{Random phase approximation}

We consider our system 
with an additional  small {\it ac} bias $\delta U$
superimposed to the source-drain bias so that
\begin{equation}
U_{SD} (t)= U_{SD} +\delta U e^{-i(\omega + i \eta) t},
\end{equation}
where $\eta \rightarrow 0$, $\eta >0$, is an
adiabatic turning-on parameter. Because of the good screening
in the contacts the applied {\it ac} bias
is assumed to lead to a dynamic potential perturbation 
$\delta \phi ({\bm r},t)
= \delta \phi ({\bm r}) \exp{[ -i(\omega + i \eta) t]}$ 
fulfilling the boundary condition
$\delta \phi (x,y,z\leq -d) = 0$ and 
$\delta \phi (x,y,z\geq d)  = \delta U$.
Then, the perturbation in the potential energy felt by electrons is
$\delta V({\bm r},t)=-e \delta \phi({\bm r})
                      \exp\biglb(-i(\omega + i \eta) t\bigrb)$ 
and the time dependent Hamiltonian becomes
\begin{equation}
\hat{H}=\hat{H}_0+\int d^3r\, \hat{\rho}({\bm r})\delta V({\bm r},t),
\label{hamilton}
\end{equation}
where $\hat{\rho}({\bm r})$ is the particle density operator.
Of course, Eq.\ (\ref{hamilton}) is only valid
if retardation effects in the quantum system can be neglected,
i. e. $\omega  \ll c/L$, where $L$ is the typical length
of the device. For a quantum device
with $L=10 \, nm$ we find $c/L \sim 10^{16}Hz$, which
is on the upper limit  of the $UV$-radiation.
The time-dependent perturbation in  the electron density
(induced density) $\delta \rho({\bm r},t)$
is calculated outside the contacts in
random phase approximation. \cite{ehrenreich,haken,hedin} 
Assuming for the planar structure
$\delta V({\bm r},t) = \delta V(z,t)$ we show
in  Appendix \ref{appendix7}  that
$ \delta \rho({\bm r},t) = \delta  \rho(z,t) = \delta \rho(z) \exp(-i\omega  t)$,
with
\begin{equation}
\delta \rho(|z| \leq d)=\int_{-d}^{d} dz' \Pi_0(z,z',\omega) \delta V(z').
\label{defrpa}
\end{equation}
Here 
$\Pi_0(z,z',\omega) = \sum_{s,s'=1}^2 \Pi^{(ss')}_0(z,z',\omega)$,
\begin{equation}
\Pi^{(ss')}_0(z,z',\omega) = 
\lim_{\eta \rightarrow 0}
\int_{V_s}^{\infty} d\epsilon \int_{V_{s'}}^{\infty} d\epsilon'
\frac{F^{(ss')}(z,z',\epsilon, \epsilon')} 
{\epsilon - \epsilon' + \hbar (\omega+ i \eta) },
\label{zpol}
\end{equation}
and
\begin{eqnarray}
F^{(ss')}(z,z',\epsilon, \epsilon') & = &
2 \frac{  m^* }{ 2\pi \beta \hbar ^2}
g_s(\epsilon) g_{s'} (\epsilon')
\ln{ \left\{  \frac{ 1 + \exp{ \biglb(\beta (\mu_s - \epsilon)\bigrb)}}
{1 + \exp{ \biglb(\beta (\mu_{s'} - \epsilon')\bigrb)} }\right\}  } 
\nonumber \\
& & \times
\Bigl( \psi^{(s)} (\epsilon,z) \Bigr)^* \psi^{(s')} (\epsilon',z)
\Bigl( \psi^{(s')} (\epsilon',z') \Bigr)^* \psi^{(s)} (\epsilon,z').
\label{zpol1}
\end{eqnarray}
Setting the integration limits in
Eq.\ (\ref{defrpa}) we assume that there is no phase coherence of the
wave functions between the contacts and the scattering area,
$\Pi^{(ss')}_0(z,|z'|>d,\omega) =0$.
The perturbation of the effective potential outside the contacts
which enters Eqs.\ (\ref{hamilton}) and (\ref{defrpa})
is determined by the Poisson equation 
\begin{equation}
\Delta \delta V(z) = - \frac{e^2 }{\kappa_s} \delta \rho(z).
\label{coulomb}
\end{equation}
In Eq.\ (\ref{coulomb}) we assume that there is
no mobile charge in the interval $-d \leq z \leq d$ other than that of the
tunneling electrons. 
The solution of Eq. (\ref{coulomb}) obeying the boundary conditions
$\delta V(z \leq -d) = 0$ and $\delta V(z\geq d) = -e \delta U$
can be written as
\begin{equation}
\delta V(z) = \delta V_{0}(z)  +
\int_{-d}^d dz' v_0(z,z') \delta \rho(z'),
\label{Vfinal}
\end{equation}
with the homogeneous solution
\begin{equation}
\delta V_{0}(|z|\le d) = -e\delta U\frac{z+d}{2d}.
\label{vhom}
\end{equation}
In Eq.\ (\ref{Vfinal}) the
symmetrical Green's function $v_0(z,z')=-(e^2/2\kappa_s)[|z-z'|+zz'/d-d]$
for the Poisson equation obeys the boundary condition\cite{sse00} 
$v_0(z = \pm d, z')= 0$. 
Using Eqs. (\ref{defrpa}) and (\ref{Vfinal}) one obtains an
integral equation for the total potential as given by
\begin{equation}
\delta V(z) = \delta V_{0}(z)  +
\int_{-d}^d dz' \int_{-d}^d dz'' v_0(z,z') \Pi_0(z',z'',\omega) \delta V(z'').
\label{final}
\end{equation}
We write down the inverse of this equation in a convenient 
discretized form
\begin{equation}
\delta {\bm V} =
\bigl( {\bm 1} -   {\bm v}_0 {\bm \Pi}_0 \bigr)^{-1} \delta {\bm V}_0,
\label{rpadi}
\end{equation}
with
$z \rightarrow z_i = - d + (i-1) \Delta z $, $i=1...N+1$, $\Delta z  = 2d/N $,
and $N \rightarrow \infty$ so that $\int_{-d}^d dz \rightarrow  \Delta z
\sum_{i=1}^{N+1} $. 
Furthermore we define 
the $N+1 \times N+1$-matrices ${\bm v}_0$ and  ${\bm \Pi}_0 (\omega)$  with
$({\bm v}_0)_{ij} = \Delta z v_0(z_i, z_j) $
and $({\bm \Pi}_0 )_{ij} = \Delta z \Pi_0(z_i,z_j,\omega)$
as well as  the vectors 
$\delta {\bm V}_0$ and $\delta {\bm V}$ with 
$(\delta {\bm V}_0)_i = \delta V_0(z_i)$ 
and $(\delta {\bm V})_i = \delta V(z_i)$.
The continuum limit of (\ref{rpadi}) 
can be regained using the von Neumann theorem,  
$\bigl( {\bm 1} -  {\bm v}_0 {\bm \Pi}_0 \bigr)^{-1} = 
\sum_{n=0}^\infty  ({\bm v}_0 {\bm \Pi}_0)^n$,
and rewriting the obtained sums as integrals. We solve 
Eq. (\ref{rpadi}) numerically.  Then after defining the 
vectors $\delta \bm{\rho}$ with $(\delta \bm{\rho})_i = \delta \rho(z_i)$
and an analogous vector $\delta {\bm j}_z$ for the z-component 
of the particle current density one obtains from discretization of
Eq.\ (\ref{defrpa})
\begin{equation}
\delta \bm{\rho} = {\bm \Pi}_0  \delta {\bm V} = {\bm \Pi}_0
\bigl( {\bm 1} -  {\bm v}_0 {\bm \Pi}_0 \bigr)^{-1} \delta {\bm V}_0,
\label{indude}
\end{equation}
and from Eq.\ (\ref{defrpacu})
\begin{equation}
\delta  {\bm j}_z  =  \tilde{\bm{\Pi}}_0 \delta {\bm V}
= \tilde{\bm{\Pi}}_0
\bigl(   {\bm 1} -   {\bm v}_0 {\bm \Pi}_0 \bigr)^{-1} \delta {\bm V}_0,
\label{cur}
\end{equation}
where $ \tilde{\Pi}_0$ is the current-density response function
defined and evaluated in Appendix \ref{appendix7} and $\tilde{\bm{\Pi}}_0$ is
the corresponding matrix obtained after discretization.

\subsubsection{AC-admittance} 

From the continuity equation $(\partial / \partial z) \delta j_z(z,t) = 
- (\partial / \partial t) \delta \rho(z,t)$
one obtains the relation
\begin{equation}
\delta {j}_z(z) = \delta j_z(-d) + 
\int_{-d}^z dz' \delta j'_z(z') = \delta j_z(-d) +
i \omega \delta Q(z) =  -\delta I / A e + i \omega \delta Q(z),
\label{conti}
\end{equation}
with $\delta j'_z(z) = (d/dz) \delta  j_z$ and 
$ \delta Q(z) =\int_{-d}^z dz' \delta \rho(z')$.
From the boundary conditions for $\delta V (z\geq d) = -e\delta U $ and
$\delta V (z \leq -d) =0$ and the continuity
of $(d/dz) \delta V (z)$
it follows that the total induced
charge $\delta Q = \delta Q(d)$ 
vanishes. One then obtains $\delta j_z(-d) = \delta j_z(d) =
-\delta I / A e$, where $\delta I$ is the induced electrical current 
provided by an external source and flowing
through the device.
The minus sign in the last step of Eq.\ (\ref{conti})
results from 
the current convention.
The complex admittance is defined as usual by
$Y = \delta I / \delta U = - A e \delta j_z(-d)/ \delta U$.
Applying
Eqs.\ (\ref{conti}), (\ref{cur}), and (\ref{vhom}) one obtains
an explicit expression for the admittance
\begin{equation}
Y = \frac{\delta I}{\delta U} = A e^2
{\bm W}_1^T  \tilde{\bm{\Pi}}_0
   \bigl({\bm 1} - {\bm v}_0 {\bm \Pi}_0 \bigr)^{-1} {\bm W}_0  ,
\label{defY}
\end{equation}
where
$({\bm W}_0)_i = z_i/ 2d + 0.5$,
and ${\bm W}_1$ is a unit vector $({\bm W}_1)_i = \delta_{1i}$.

\section{Quantum capacitor in random phase approximation} 

\subsection{General results}

We define a  quantum capacitor through two conditions: First,
there is basically no dc currents traversing the structure,
\begin{equation}
\psi^{(1)} (\epsilon, z \geq d) =0, \quad 
\psi^{(2)} (\epsilon, z \leq -d) = 0.
\label{idealcon}
\end{equation}
Second, the overlap of the
right-incident and the left-incident scattering functions can be neglected,
\begin{equation}
\psi^{(1)} (\epsilon,z) \psi^{(2)} (\epsilon',z) \sim 0,
\end{equation}
for all $\epsilon,\epsilon'$ and $z\in [-d,d]$.
We then find from Eqs.\ (\ref{zpoltil}) and (\ref{zpol})
$\tilde {\bm \Pi}_0^{(12)} =  \tilde{\bm \Pi}_0^{(21)} =
 {\bm \Pi}_0^{(12)} =  {\bm \Pi}_0^{(21)} =0$,
and the induced current can be split into two independent
parts, $\delta j_z = \delta j^{(1)}_z + \delta j^{(2)}_z$, 
with
\begin{equation}
\delta j^{(s)}_z(z) = 
\int_{-d}^{d} dz' \tilde{\Pi}^{(ss)}_0(z,z',\omega) \delta V(z').
\label{sepa}
\end{equation}
Since $\delta j^{(s)}_z$
results exclusively from the source-incident scattering states for $s=1$
or exclusively from the drain-incident scattering states for $s=2$ 
we can write for each component a separate continuity equation,
\begin{equation}
- i \omega \delta \rho^{(s)} (z) + \frac{d}{dz} \delta j^{(s)}_z(z) =0,
\label{sepcon}
\end{equation}
with
\begin{equation}
 \delta \rho^{(s)} (z) = 
 \int_{-d}^{d} dz' \Pi_0^{(ss)}(z,z',\omega) \delta V(z').
\end{equation}
Integrating Eq.\ (\ref{sepcon}) one obtains 
using Eq. (\ref{idealcon}) 
under another form, i.e. $\delta j_z^{(1)}(d)=\delta j_z^{(2)}(-d)=0$,
\begin{equation}
\frac{\delta I }{eA}  =  - \delta j_z^{(1)}(-d) = i\omega
\int_{-d}^d dz \int_{-d}^d dz' \Pi_0^{(11)} (z,z',\omega) \delta V(z').
\end{equation}
With the definition of the admittance and Eqs.\ (\ref{rpadi}) and (\ref{vhom})
the equation (\ref{defY}) reduces after discretization to  
\begin{equation}
Y =   - e^2 A i \omega \Delta z  {\bm W}^T_2  {\bm \Pi}^{(11)}_0
  \bigl({\bm 1} - {\bm v}_0 {\bm \Pi}_0\bigr)^{-1} {\bm W}_0,
       \label{zcae}
\end{equation}
where $ \tilde{\bm{\Pi}}_0$ is eliminated.
In Eq.\ (\ref{zcae}
we define the $N+1 \times N+1$-matrix  $({\bm \Pi}^{(ss)}_0)_{ij} =
\Delta z \Pi_0^{(ss)}(z_i,z_j,\omega)$ and the vector
$({\bm W}_2)_i = 1 $.

In Appendix \ref{lowfreqexp} it is shown that for small frequencies an expansion
\begin{equation}
\Pi_0^{(ss)}(z,z',\omega)  = P^{(s)}_0 (z,z')
+ i \omega P^{(s)}_1 (z,z')
\label{expa}
\end{equation}
can be derived with real functions $P_0^{(s)}(z,z')$ and $P_1^{(s)}(z,z')$. 
Inserting this expansion in Eq.\ (\ref{zcae}) one obtains 
a low-frequency expansion for the admittance of a quantum capacitor
as
\begin{equation}
Y \approx   -  i \omega (Y_1 +  i \omega Y_2).
\label{lowfr}
\end{equation}
Here the leading order  coefficient
\begin{equation}
Y_1 =  e^2 A \Delta z
{\bm W}^T_2 {\bm P}^{(1)}_0 \bigl(1 - {\bm v}_0 {\bm P}_0\bigr)^{-1}
{\bm W}_0,
\end{equation}
and the first correction
\begin{equation}
Y_2 =  e^2 A \Delta z {\bm W}^T_2
\left[ {\bm P}^{(1)}_1  \bigl(1 - {\bm v}_0 {\bm P}_0 \bigr)^{-1} +
        {\bm P}^{(1)}_0 \bigl(1 - {\bm v}_0 {\bm P}_0 \bigr)^{-2}
	          {\bm v}_0 {\bm P}_1 \right]
		  {\bm W}_0,
\end{equation}
are real,
where ${\bm P}_{0} = {\bm P}^{(1)}_{0} + {\bm P}^{(2)}_{0}$,
${\bm P}_{1} = {\bm P}^{(1)}_{1} + {\bm P}^{(2)}_{1} $,
$({\bm P}^{(s)}_0)_{ij} =
\Delta z P_0^{(s)}(z_i,z_j,\omega)$, and $({\bm P}^{(s)}_1)_{ij} =
\Delta z P_1^{(s)}(z_i,z_j,\omega)$ .

\subsection{Numerical results for MIS-type nanostructure}

\begin{figure}[ht]
\noindent\includegraphics*[width=3.25in]{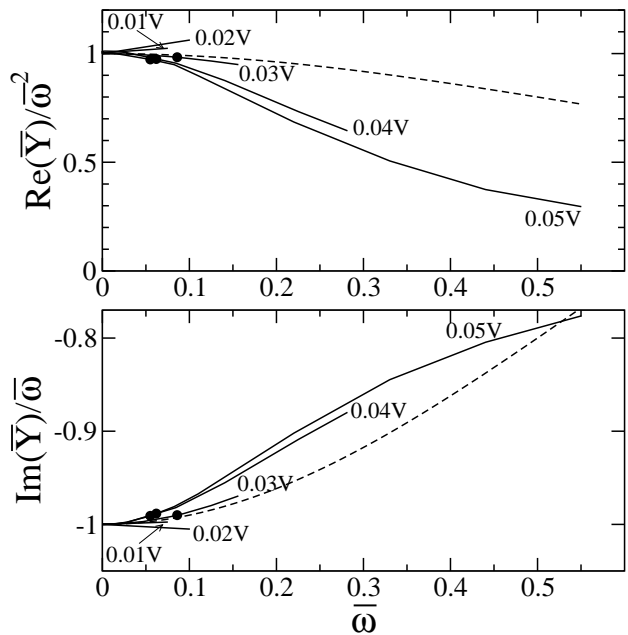}
\caption{Quantum admittance for MIS-type nanostructure, 
in a normalized representation, for different
static biases $U_{SD}$, parameters corresponding to Fig.\ 1 of 
Ref. \ [\onlinecite{prb02}].
Dashed line represents $Y_{sg}(\bar{\omega})$ 
for a classical RC-circuit with frequency-independent elements.  
The symbols represent the critical frequencies\protect\cite{emrs05} 
up to which the approximation of the quantum result (i.e. Eq. (\ref{zcae})) 
with $Y_{sg}$
can be  considered satisfactory.}
\label{fig2}
\end{figure}

As a test structure for our theory  we take a planar
MIS (metal insulator semiconductor)-type GaAs/Al$_x$Ga$_{1-x}$As 
heterostructure with a near back gate. This structure has been analyzed in
experiments,\cite{dolgopolov} the stationary system has been
described  theoretically within Hartree approximation \cite{prb02}
and first calculations for the dynamic behavior are presented in 
Refs.\ [\onlinecite{emrs03,emrs05}]. 
We now calculate the quantum admittance $Y(\omega)$ according to 
Eq.\ (\ref{zcae}) up to  frequencies of 100GHz.
Using Eq.\ (\ref{lowfr}) one can extract
from the numerically (or experimentally) given data the parameters
$Y_1 = - \lim_{\omega \rightarrow 0} Im[ Y(\omega)]/\omega$ and
$Y_2 = \lim_{\omega \rightarrow 0} Re[Y(\omega)]/\omega^2$. It is then possible
to recast  Eq.\ (\ref{lowfr}) in a normalized form
\begin{equation}
\bar{Y} = -i \bar{\omega} ( 1 + i \bar \omega),
\label{lownorm}
\end{equation}
with $\bar{\omega} = \omega/  \omega_0$, $\omega_0 = Y_1 / Y_2$
and $\bar{Y} = Y  Y_2 / Y_1^2$.
This normalization allows us to collapse the calculated quantum admittance 
at all considered source-drain voltages into one graph which is presented
in Fig. \ref{fig2}. 
 It is immediately seen that the expansion 
in Eq. (\ref{lownorm}) only holds for $\bar{\omega} \rightarrow 0$.
For finite frequencies there are significant deviations from the value $1$
for $Re(\bar{Y})/\bar{\omega}^2$ and 
from the value $-1$ for $Im(\bar{Y})/\bar{\omega}$.
To discuss these deviations we compare with the admittance
$Y_{sg}$ of a classical equivalent
circuit consisting of a frequency-independent resistance $R$ and 
a frequency-independent capacitor $C$ 
in series.\cite{buettiker93c,buettiker96c,buttiker98} 
An inspection of 
the admittance $Y_{sg}(\omega \rightarrow 0)$ of this circuit
yields $Y_1 = C$ and $Y_2 = R C^2$. These formulae can be regarded as 
quantum mechanical
expressions for the elements of the equivalent circuit. In the normalized form 
one then obtains 
\begin{equation}
\bar{Y}_{sg}(\bar{\omega}) =\frac{\bar{\omega}^2}{1+\bar{\omega}^2 }
              -i\frac{\bar{\omega} }{1+\bar{\omega}}.
\end{equation}
It is seen from Fig. \ref{fig2} that $Y_{sg}(\bar{\omega})$ 
generally fails to describe the numerical admittance.
The numerical results show for increasing source-drain voltages
a systematic enhancement of
the decrease in $Re(\bar{Y})$
and the increase in $Im(\bar{Y})$ 
as the frequency is increased.
This finding does not result in the
classical equivalent circuit.

\begin{figure}[ht]
\noindent\includegraphics*[width=3.25in]{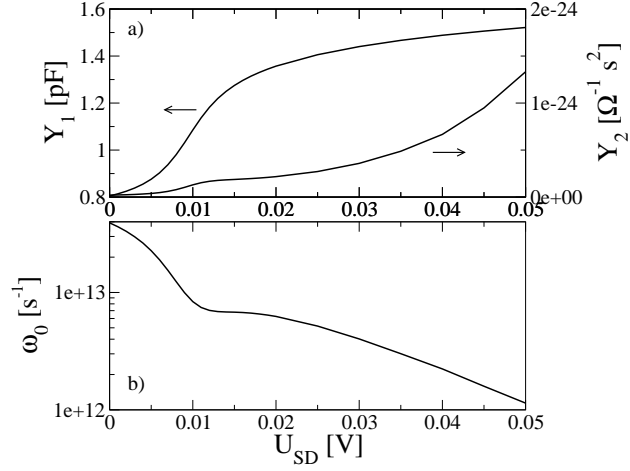}
\caption{a) $Y_1$ and $Y_2$ as function of the working point $U_{SD}$.
b) The scaling frequency $\omega_0$ vs. $U_{SD}$.
}
\label{fig3}
\end{figure}

In Fig. \ref{fig3} (a) we represent the dependence of the coefficients $Y_1$
and $Y_2$ on the working point bias $U_{SD}$. The coefficient
$Y_1$ is essentially identical with the low frequency limit
of the dynamic capacitance plotted in Fig.\ 3  of 
Ref. \ [\onlinecite{emrs05}].
As demonstrated in Ref. \ [\onlinecite{emrs05}] this step in $Y_1$
is in good agreement with
a step in the experimental capacitance curve 
which is caused by the formation of a 
two-dimensional electron gas within the quantum capacitor. 
The coefficient $Y_2$ shows a general increase with increasing bias.
As a characteristic feature it is seen that the step in the capacitance of $Y_1$
is accompanied by a small hump in $Y_2$. 
An inspection of the scaling frequency $\omega_0$ 
(Fig. \ref{fig3} (b)) reveals a corresponding downward hump.

\section{Conclusions}
We present a quantum mechanical model  to calculate the admittance of
effectively one-dimensional open quantum systems
in random phase approximation.
Explicit RPA-expressions
for the quantum admittance of a general two-terminal system
are derived.
In the case of a quantum capacitor 
a small-frequency expansion can be obtained
which lends itself to an experimental testing of the theory.
A comparison of the low-frequency expansion
with the complete RPA-expression shows that for a
quantum capacitor a simple classical equivalent circuit with
frequency-independent elements does not
describe satisfactorily the quantum-admittance with increasing the
frequency.

\begin{acknowledgments} 

It is a pleasure for us to acknowledge intense discussions with D. Robaschik.

We gratefully acknowledge support by the HWP grant of the Federal
Government of Germany and the State of Brandenburg.
Two of us (P.N.R and E.R.R.) also acknowledge partial support from the
CEEX project D-11-45 from Romanian Ministry of Education and Research.
\end{acknowledgments}

\appendix

\section{Statistical operator for
the stationary system with finite source-drain bias}
\label{appendix6}

\subsection{Scattering states as a complete orthonormal single particle basis}

As shown in Ref. \ [\onlinecite{roxana}] 
the scattering states $\psi^{(s)} (\epsilon,z)$
constitute a complete orthonormal system, i. e.
\begin{equation}
\sum_{s=1,2} \int_{V_s}^{\infty} d\epsilon \; g_s(\epsilon)
\left( \psi^{(s)}(\epsilon,z) \right)^{*} \psi^{(s)}(\epsilon,z')
= \delta(z - z'),
\label{complet}
\end{equation}
and
\begin{equation}
\int_{-\infty}^{\infty} d z
\left( \psi^{(s)}(\epsilon,z) \right)^{*} \psi^{(s')}(\epsilon',z)
= \theta(\epsilon-V_s) \delta_{ss'} \delta(\epsilon-\epsilon')/g_s(\epsilon).
\label{ortho}
\end{equation}
To eliminate  the weight function $g_s(\epsilon)$ we substitute
for a given $s$ in Eq.\ (\ref{complet}) 
$\epsilon = \hbar^2 k_s^2 /(2m^*) + V_s \equiv \epsilon_s (k_s)$ (see
Eq.\ (\ref{6k})).
One obtains 
\begin{equation}
\sum_{s=1,2} \int_{0}^{\infty} d k_s
\left( \psi^{(s)}\biglb( \epsilon_s(k_s),z\bigrb) \right)^{*} 
       \psi^{(s)}\biglb( \epsilon_s(k_s),z'\bigrb)
= \delta(z - z').
\label{completek}
\end{equation}
According to  Eq.\ (\ref{6k}) we write in
Eq.\ (\ref{ortho}) 
$\epsilon' = \hbar^2 (k'_{s'})^2 /(2m^*) + V_{s'} \equiv \epsilon_{s'} (k'_{s'})$
and using the identity 
$\delta_{ss'}\delta(\epsilon-\epsilon')/g_s(\epsilon) = \delta_{ss'}
\delta(k_s-k_s')$ it results that
\begin{equation}
\int_{-\infty}^{\infty} d z
\left( \psi^{(s)}\biglb( \epsilon_s(k_s),z\bigrb) \right)^{*} 
\psi^{(s')}\biglb( \epsilon_{s'}(k'_{s'}),z\bigrb)
= \theta(k_s) \delta_{ss'} \delta(k_s-k_s').
\label{orthok}
\end{equation}
Since later we want to work in a number representation
we introduce a $k$-space
discretization $k_s \rightarrow k_j = j \Delta k$, $j=0,1,2....$,
so that $\delta(k_s-k_s') \rightarrow \delta_{jj'}/\Delta k$. 
Furthermore, we define
$\psi_{sj} (z) = \sqrt{\Delta k}\, \psi^{(s)}\biglb(\epsilon_s(k_{j}),z\bigrb)$.
With this definition 
an  explicit asymptotic form of the source-incident
scattering wave functions ($s=1$) follows
from Eq.\ (\ref{6psi-out-1})  as given by
\begin{equation}
 \psi_{1j} (z) =\sqrt{ \frac{\Delta k}{2\pi} }  
                      \theta\biglb( \epsilon_{1} (k_j) -V_1) \bigrb)
 \left\{
 \begin{array}{ll}
  \exp{\biglb(i k_j (z+d)\bigrb)} 
+ {\bm S}_{11}\biglb(\epsilon_{1} (k_{j})\bigrb) 
  \exp{\biglb(-i k_j (z+d)\bigrb)},
 & \mbox{~~for~~} z \le -d \\
 & \\
 {\bm S}_{21}\biglb(\epsilon_{1}(k_{j}) \bigrb)
\exp{\Biglb(i \sqrt{k_j^2+\frac{2m^*}{\hbar^2}(V_1-V_2)}(z-d)\Bigrb)},
 & \mbox{~~for~~} z \ge d.
 \end{array}
\right. 
\end{equation}
For the drain-incident scattering functions ($s=2$) we find the asymptotics
\begin{equation}
\psi_{2j}(z) = \sqrt{ \frac{\Delta k}{2\pi} } 
                 \theta\biglb(\epsilon_{2}(k_j) -V_2\bigrb)
                         \left\{
                         \begin{array}{ll}
                   {\bm S}_{12}\biglb(\epsilon_{2}(k_{j})\bigrb) 
\exp{\Biglb(-i \sqrt{k_j^2+\frac{2m^*}{\hbar^2}(V_2-V_1)} (z+d)\Bigrb)},
                                & \mbox{~~for~~} z \le -d \\
                                & \\
                                 \exp{\biglb(-i k_j (z-d)\bigrb)}
                      +{\bm S}_{22}\biglb(\epsilon_{2}(k_{j})\bigrb)
                                  \exp{\biglb(i k_j (z-d)\bigrb)},
                                & \mbox{~~for~~} z \ge d.
                        \end{array}
                        \right. 
\end{equation}

After discretization we thus write
for (\ref{completek}) 
\begin{equation}
\sum_{s=1}^2 \sum_{j=0}^{\infty} 
\psi_{sj}^{*}(z) \psi_{sj}(z')
= \delta(z - z'),
\label{completedis}
\end{equation}
and for (\ref{orthok})
\begin{equation}
\int_{-\infty}^{\infty} d z \, 
\psi_{sj}^{*}(z) \psi_{s'j'} (z)
= \delta_{ss'} \delta_{jj'}.
\label{orthodis}
\end{equation}
In addition we introduce a complete orthonormal basis system 
$ \phi_{\nu}({\bm r}_{\perp})$ for the
square-integrable function in ${\bm R}^2$,
$\int d {\bm r}_{\perp} \, \phi^*_{\nu}({\bm r}_{\perp})
      \phi_{\nu'}({\bm r}_{\perp})
 = \delta_{\nu \nu'}$, and
$\sum_{\nu} \phi^*_{\nu}({\bm r}_{\perp})
  \phi_{\nu}({\bm r'}_{\perp})
   =  \delta({\bm r}_{\perp}-{\bm r'}_{\perp})$. 
They are usually solutions of the time-independent Schr\"odinger equation
in the lateral directions, 
$[-\hbar^2/(2m^*)\Delta_\perp+V_\perp({\bm r}_\perp)-E_\perp^\nu]
\phi_{\nu}({\bm r}_{\perp})=0$.
Then  a complete orthonormal basis for the Hilbert space of the 
single particle quantum states is given by
\begin{equation}
\varphi_{\alpha} ({\bm r}) = < {\bm r}| \varphi_{\alpha} >  
= \psi_{sj} (z) \phi_{\nu}({\bm r}_{\perp}),
\end{equation}
where $\alpha$ is the index-triple $(sj\nu)$.
We find the usual completeness relation 
$\sum_{\alpha} \varphi^{*}_{\alpha} ({\bm r})  \varphi_{\alpha} ({\bm r'})
=  \delta({\bm r}-{\bm r'})$
and the   orthonormality relation is given by
$
\int d^3 r  \varphi^{*}_{\alpha} ({\bm r}) \varphi_{\alpha'} ({\bm r})
= \delta_{\alpha \alpha'}$.

\subsection{Definition of the statistical operator}

Because the scattering states
$|\varphi_{\alpha}>$ constitute a discrete and complete orthonormal basis
it is possible to introduce the creation- and annihilation
operator $\hat{c}_{\alpha}^{\dagger}$ and  $\hat{c}_{\alpha}$, respectively,
with the usual anticommutation relations
$\{ \hat{c}_{\alpha} \, , \, \hat{c}_{\alpha'}^{\dagger} \}
=    \delta_{\alpha, \alpha'}$, 
$\{ \hat{c}_{\alpha}, \, \hat{c}_{\alpha'} \} = 0$, and
$\{ \hat{c}^{\dagger}_{\alpha}, \, \hat{c}^{\dagger}_{\alpha'} \} =   0$.
Based on the  anticommutation relations one can formulate
a particle number representation which will be described in the following.

According to standard theory\cite{nolting7} the field operators are given by
$\hat{\Psi}  ({\bm r})  = 
\sum_{\alpha}  \varphi_{\alpha} ({\bm r}) \hat{c}_{\alpha} $
and $ \hat{\Psi}^{\dagger}  ({\bm r})  =
\sum_{\alpha} \varphi_{\alpha}^* ({\bm r})  \hat{c}_{\alpha}^{\dagger} $.
The many-particle Hamiltonian of the stationary 
electron system can be written as
\begin{eqnarray}
\hat H_0 &=&  \int d^3 r \; \hat \Psi^{\dagger}({\bm r})
                           \left[ -\frac{\hbar^2}{2 m^*} \Delta
                                 +V_{\perp}({\bm r}_{\perp})
                                 +V(z)
                           \right]
                         \hat \Psi({\bm r}),
        \nonumber \\  
& = & \sum_{\alpha}  
              E_{\alpha} \hat{c}_{\alpha}^{\dagger} \hat{c}_{\alpha},
\label{Ham}
\end{eqnarray}
where $ E_{\alpha} = E_{\perp}^{\nu} + \epsilon_s(k_i)$
is the energy of the single particle state $| \varphi_{\alpha} >$.
Here we use the notation $\hat H_0$ since in Sect. \ref{harmpert} 
we consider a time dependent
perturbation of this Hamiltonian.
We now use as a  statistical operator for the stationary system as given by 
\begin{equation}
\hat{\rho}_0=\frac{1}{Z_0}
             \exp[-\beta (\hat{H}_0-\mu_1\hat{N_1}-\mu_2\hat{N_2})],
\label{stat}
\end{equation}
with the chemical potentials $\mu_1$ and $\mu_2$ of source- and drain contact,
respectively, the particle number operators 
$\hat{N_s} =  \sum_{i\nu}\hat{c}_{si\nu}^{\dagger} \hat{c}_{si\nu}$
with $\hat{N} = \hat{N}_1 + \hat{N}_2$, and
\begin{equation}
Z_0=\mbox{Tr}\{\exp[-\beta (\hat{H}_0-\mu_1\hat{N_1}-\mu_2\hat{N_2} ]\}.
\end{equation}
The trace is done over all states of a Fock-space basis,
constructed with the help of the single particle scattering states 
$|\varphi_{\alpha}>$. 
The trace is easy to write in the occupation number representation
\begin{eqnarray}
Z_0 &=& \sum_{N=0}^\infty \sum_{\{n_{\alpha}\}}
  <N; \dots n_{\alpha} \dots |
\exp[-\beta \sum_{\alpha'}(E_{\alpha'}-\mu_{s'})
 \hat{c}_{\alpha'}^{\dagger} \hat{c}_{\alpha'} ] | N; \dots n_{\alpha} \dots> 
\nonumber \\
& = & 
\prod_{\alpha} \left(1+\exp\biglb(-\beta (E_{\alpha}-\mu_s)\bigrb) \right).
\label{z0}
\end{eqnarray}
The occupation number $n_{\alpha}$ of the single particle state 
$|\varphi_{\alpha}>$ for electrons is 0 or 1.
It is straightforward to show that the statistical operator in
Eq.\ (\ref{stat}) is stationary,
i. e. $[\hat{H}_0, \hat{\rho}_0  ] =0$ and that it fulfills
$\mbox{Tr}\{ \hat{\rho}_0 \} =1$.

\subsection{Expectation values}

Using Eq.\ (\ref{z0}) one finds 
for the equilibrium mean value $\langle n_{\alpha}\rangle $ 
of the particle number
operator of a single particle state $|\varphi_{\alpha}>$
\begin{equation}
\mbox{Tr}\{\hat{\rho}_0 \hat{c}^\dagger_{\alpha} \hat{c}_{\alpha'} \}  = 
\delta_{\alpha \alpha'} \langle n_{\alpha} \rangle  = \delta_{\alpha \alpha'}
\frac{\exp\biglb(-\beta(E_{\alpha}-\mu_s)\bigrb)}
     {1+\exp\biglb(-\beta(E_{\alpha}-\mu_s)\bigrb)}
= \delta_{\alpha \alpha'} f_{FD}(E_{\alpha}-\mu_s),
\label{FD}
\end{equation} 
with the Fermi-Dirac distribution function
\begin{equation}
f_{FD}(E_{\alpha}-\mu_s) = \frac{1}
                       {1+ \exp {\biglb( \beta (E_{\alpha}- \mu_s ) \bigrb)}}.
\label{FDdef}
\end{equation}
In agreement with the Landauer-B\"uttiker 
formalism it then results for the mean value of the particle density 
operator, 
\begin{eqnarray}
\rho({\bm r}) &=&2 \mbox{Tr}\left\{ \hat \rho_0 \,
                               \hat \Psi^{\dagger}({\bm r}) \hat \Psi({\bm r})
                        \right\} 
  = 2 \sum_{\alpha} f_{FD}(E_{\alpha}-\mu_s) |\varphi_{\alpha}({\bm r})|^2 
\nonumber \\
  & = & 2\sum_{s\nu}
  \int_{V_s}^{\infty} d \epsilon \; g_s(\epsilon)
   f_{FD}( E_{\perp}^{\nu} +\epsilon-\mu_s)
   \left| \phi_{\nu}({\bm r}_{\perp}) \right|^2
    \left| \psi^{(s)}(\epsilon,z) \right|^2,
     \label{qq-2}
\end{eqnarray}
where we establish the continuous limit by replacing
\begin{equation}
 \Delta k \sum_{\alpha}  
\rightarrow \sum_{s\nu} \int_0^\infty dk_s
\rightarrow 
 \sum_{s\nu} \int_{V_s}^{\infty} d \epsilon \; g_s(\epsilon) 
\label{contialpha}
\end{equation}
and the factor two comes from the spin degeneracy.
Considering the ansatz (\ref{6ansatz}) 
one can perform the $\nu$ summation 
in Eq. (\ref{qq-2}) obtaining Eq.(\ref{6qq-2}). 

\section{Harmonic perturbation}
\label{appendix7}

In this Appendix we describe our approach to calculate 
dynamic linear-response properties of open quantum systems 
which are defined in Sect.\ref{stat_LB}. 
Starting with Eq.\ (\ref{hamilton})
we determine the density matrix for the perturbed system 
using the von Neumann equation with $\hat{H}_0$ describing the stationary
open system, Eq. (\ref{Ham}).
In linear approximation one finds \cite{czy}
\begin{equation}
\delta  \rho({\bm r}, \omega)  
= \int d^3 r'\, \Pi_0({\vrv},{\vrv'},\omega)
\delta V({\bm r'},\omega) ,
\label{WWmvdB}
\end{equation}
where the density-density correlation function 
(irreducible polarization) is given by
\begin{equation}
\Pi_0({\vrv},{\vrv'},\omega) =
\frac{i}{\hbar} \int_0^{\infty} d\tau \, \exp\biglb(i(\omega+i\eta) \tau\bigrb)
\mv{[\hat{\rho}_I({\bm r},\tau), \hat{\rho} ({\bm r}')]}_{0} .
\label{WWchi}
\end{equation}
The index $0$  means the thermodynamic expectation value
with respect to the statistical operator $\hat{\rho}_0$ 
as given by  Eq.\ (\ref{stat})
and the index $I$ means the operator in the interaction picture.
The single particle density operators $\hat{\rho}$ 
are now written in the second quantization using the field operators
of the scattering states defined in Appendix \ref{appendix6} so that
\begin{equation}
\hat{\rho}_I({\bm r},\tau) 
   = \sum_{\alpha, \alpha'}
     \varphi_{\alpha}^* ({\bm r})  \varphi_{\alpha'} ({\bm r})
     \exp{ [i \hat{H}_0\tau/\hbar]}
     \hat{c}_{\alpha}^\dagger \hat{c}_{\alpha'}
     \exp{ [- i \hat{H}_0\tau/\hbar]} ,
\end{equation}
with $\alpha \equiv (sj\nu)$ and 
$\hat{\rho}({\bm r}) = \hat{\rho}_I({\bm r},\tau =0)$.
In a standard way one uses the anticommutation relations 
for the $\hat{c}_\alpha^\dagger$ and $\hat{c}_\alpha$ 
to calculate the commutator in Eq. (\ref{WWchi}) and obtains
\begin{equation}
\Pi_0({\vrv},{\vrv'},\omega) = 2 \lim_{\eta \rightarrow 0}
\sum_{\alpha,\alpha'}
\frac{f_{FD} (E_\alpha - \mu_s)  -f_{FD} (E_{\alpha'} - \mu_{s'})}
{E_\alpha - E_{\alpha'} + \hbar(\omega+i \eta)}
\varphi_\alpha^*({\bm r}) \varphi_{\alpha'}({\bm r})
\varphi_{\alpha'}^*({\bm r'}) \varphi_\alpha({\bm r'}),
\label{pol_rs}
\end{equation}
where we have used Eq.\ (\ref{FD}) for the expectation values.
In difference to the usual expression
for $\Pi_0$ the Fermi-Dirac occupation functions
$f_{FD}$ may contain
different chemical potentials, either  that of the
source contact for $s=1$ or that of the drain contact for $s=2$.
Furthermore, the single particle energies $E_\alpha$ are 
given
in  Eq.\ (\ref{Ham}) and
the overall factor of two in Eq.\ (\ref{pol_rs}) accounts for
the spin degree of freedom. 

In the considered planar structure
the potential perturbation depends only on
$z$, $\delta V({\vrv'}) =  \delta V(z)$. 
In order to develop further Eq. (\ref{pol_rs}), 
we take the continuous limit [Eq. (\ref{contialpha})] and 
using Eqs. (\ref{totalE}) and (\ref{6ansatz})
one can first integrate over ${\bm r}_\perp$ obtaining  
$\delta({\bm k}_\perp - {\bm k'}_\perp)$ and after that 
sum over ${\bm k}_\perp$.
As a result 
a $z$-dependent density modulation $\delta \rho({\vrv}) = \delta \rho(z)$
is obtained
which is related to the potential modulation by Eq.\ (\ref{defrpa}), 
and the polarization $\Pi_0(z,z',\omega)$ is given by Eq.\ (\ref{zpol}).

Evaluating in the same procedure
the perturbation of the
expectation value of the
z-component of particle current density 
$\hat{j}_z({\bm r})=(\hbar/m^*)
                          {\mbox Im}\left(\hat{\Psi}^{\dagger}({\bm r})
                                \nabla_z \hat{\Psi}({\bm r}) \right)$ 
one obtains 
\begin{equation}
\delta j_z(|z| \leq d)=\int_{-d}^{d} dz' 
\tilde{\Pi}_0(z,z',\omega) \delta V(z').
\label{defrpacu}
\end{equation}
The current-density response function for a planar structure has the form
$\tilde{\Pi}_0(z,z',\omega) = \sum_{ss'} \tilde{\Pi}^{(ss')}_0(z,z',\omega)$
with
\begin{eqnarray}
\tilde{\Pi}^{(ss')}_0(z,z',\omega) &=&
\lim_{\eta \rightarrow 0}
\int_{V_s}^{\infty} d\epsilon \int_{V_{s'}}^{\infty} d\epsilon'
\frac{\tilde{F}^{(ss')}(z,z',\epsilon,\epsilon')}
     {\epsilon - \epsilon' + \hbar (\omega+ i \eta) } 
\label{zpoltil}
\end{eqnarray}
and
\begin{eqnarray}
\tilde{F}^{(ss')}(z,z',\epsilon,\epsilon') &=& 
 2 \frac{ m^* }{2\pi \beta \hbar ^2}
g_s(\epsilon) g_{s'} (\epsilon')
\ln{ \left\{ \frac { 1 + \exp{ [\beta (\mu_s - \epsilon)]}} 
{1 + \exp{ [\beta (\mu_{s'} - \epsilon')] } }\right\}  } \nonumber \\
& & \times \frac{\hbar}{2im^*} \left[\Bigl(\psi^{(s)}(\epsilon,z)\Bigr)^* 
                    \frac{d}{dz} \psi^{(s')} (\epsilon',z)
                   -\psi^{(s')} (\epsilon',z)
                    \frac{d}{dz} \Bigl(\psi^{(s)}(\epsilon,z)\Bigr)^*
              \right] \nonumber \\
& & \times 
\Bigl(\psi^{(s')} (\epsilon',z')\Bigr)^* \psi^{(s)} (\epsilon,z').
\label{zpoltil1}
\end{eqnarray}

\section{Low frequency expansion of the irreducible polarization}
\label{lowfreqexp}

\begin{figure}[t]
\noindent\includegraphics[width=2.5in]{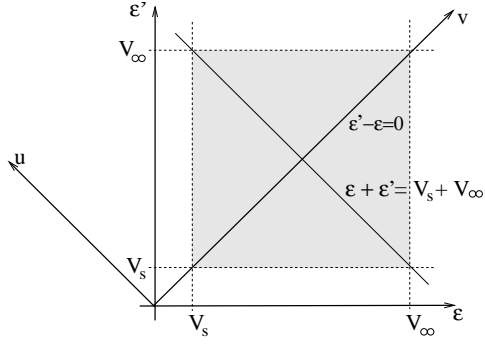}
\caption
{Transformation of the two-dimensional integral (\protect\ref{2dintr}).
The gray area illustrates the integration domain.}
\label{fig4}
\end{figure}

For small frequencies we expand in Eq.\ (\ref{zpol}) for finite $\eta$
\begin{equation}
\frac{1}{\epsilon - \epsilon' + \hbar (\omega+ i \eta) }
\approx \frac{1}{\epsilon - \epsilon' +  i  \hbar \eta } 
      - \frac{\hbar \omega}{( \epsilon - \epsilon' + i \hbar \eta)^2 },
\end{equation}
finding
\begin{equation}
P_0^{(s)} (z,z')= \lim_{\eta \rightarrow 0}
\int_{V_s}^{\infty} d\epsilon \int_{V_{s}}^{\infty} d\epsilon'
\frac{F^{(ss)}(z,z',\epsilon, \epsilon')}
{\epsilon - \epsilon' + i \hbar \eta },
\label{p0s}
\end{equation}
and
\begin{equation}
P_1^{(s)}(z,z') = -\frac{\hbar}{i} \lim_{\eta \rightarrow 0}
\int_{V_s}^{\infty} d\epsilon \int_{V_{s}}^{\infty} d\epsilon'
\frac{F^{(ss)}(z,z',\epsilon, \epsilon')}
{(\epsilon - \epsilon' + i \hbar  \eta)^2 }.
\label{p1s}
\end{equation}
We introduce a transformation
$v = \epsilon + \epsilon'$ and $u = \epsilon' - \epsilon$ so that
for a general function $f(\epsilon, \epsilon')$
\begin{equation}
\int_{V_s}^{V_\infty} d \epsilon \int_{V_s}^{V_\infty} d \epsilon' 
f(\epsilon, \epsilon')
=
\frac{1}{2} \int_{2V_s}^{2 V_\infty} dv  \int_{-u_0 (v)}^{u_0 (v)} du
f\biglb( \epsilon(v,u) , \epsilon'(v,u) \bigrb),
\label{2dintr}
\end{equation}
where we introduce a cut-off
energy $V_\infty \rightarrow \infty$. Furthermore, as illustrated in 
Fig.\ \ref{fig4},
$u_0 (v< V_s +V_\infty ) = v - 2 V_s$ and $u_0 (v> V_s +V_\infty ) = 
2 V_\infty -v$.
For fixed $z,z'$  we write 
$F^{(ss)}(z,z',\epsilon, \epsilon') =  \alpha \chi(v,u)$
with $\alpha = 2m^*/2\pi \beta \hbar^2$ and
\begin{equation}
\chi(v,u) = M\left( \frac{v-u}{2} \right) M^*\left( \frac{v+u}{2} \right)
\left[ N\left( \frac{v-u}{2} \right) - N\left( \frac{v+u}{2} \right) \right],
\end{equation}
where
\begin{equation}
M(v) = g_s (v) \Bigl(\psi^{(s)}(v,z)\Bigr)^* \psi^{(s)}(v,z')
\end{equation}
and  
\begin{equation}
N(v) =  \ln { \left\{ 1+ \exp {\biglb( \beta (\mu_s - v) \bigrb)} \right\} }.
\end{equation}
It is easy to see that $\chi(v,u) = \chi_1(v,u) + i \chi_2(v,u)
= - \chi^*(v,-u)$ so that $\chi_1(v,u) = - \chi_1(v,-u)$ and 
$\chi_2(v,u) = \chi_2(v,-u)$. Furthermore, since $\chi(v,0) =0$
one finds for small $|u|$  the expansion
\begin{equation}
\chi(v,u \rightarrow 0) \approx u   \chi_u(v,0)
\label{expachi}
\end{equation}
where we obtain a real function for the partial derivative with respect to $u$
at $u=0$
\begin{equation}
 \chi_u(v,0) = \beta \left|  M\left(\frac{v}{2}\right)  \right|^2 
f_{FD}\left( \frac{v}{2} - \mu_s \right),
\label{chip}
\end{equation}
with the Fermi-Dirac distribution function given by Eq. (\ref{FDdef}).
This means that in the expansion (\ref{expachi}), the leading term
in the real part is linear in $u$ while the leading term in 
imaginary part is parabolic in $u$.
Writing
$ \lim_{\eta \rightarrow 0} (\epsilon - \epsilon' + i \hbar \eta)^{-1}=
-PV(1/u) - i \pi \delta (u)$, 
where $\mbox{PV}$ denotes the Cauchy principal value 
and using the symmetry properties of the functions $\chi_1$ and $\chi_2$, 
one obtains 
\begin{equation}
P_0^{(s)}(z,z') =  - \frac{\alpha}{2} 
\int_{2V_s}^{2 V_\infty} dv  \int_{-u_0 (v)}^{u_0 (v)} du
\frac{\chi_1(v,u)}{u} 
= \int_{V_s}^{V_\infty} d \epsilon \int_{V_s}^{V_\infty} d \epsilon'
\frac{F_1^{(ss)}(z,z',\epsilon, \epsilon')}{\epsilon - \epsilon'},
\label{pioe}
\end{equation}
where $F^{(ss)}(z,z',\epsilon, \epsilon') = F_1^{(ss)}(z,z',\epsilon, \epsilon')
+ i F_2^{(ss)}(z,z',\epsilon, \epsilon')$. In Eq.\ (\ref{pioe}) we omitted the
principal value operation because one obtains in $u=0$ a regular integrand
due to the expansion in Eq.\ (\ref{expachi}).
Writing \cite{gelfand}
$ \lim_{\eta \rightarrow 0} (\epsilon - \epsilon' + i \hbar \eta)^{-2}=
 PV(1/u^2) - i \pi (d/du) \delta (u)$ it follows from
Eq.\ (\ref{p1s}) that
\begin{eqnarray}
P_1^{(s)}(z,z') &=&  -\frac{\alpha\hbar}{2}
\int_{2V_s}^{2 V_\infty} dv  \int_{-u_0 (v)}^{u_0 (v)} du
\frac{\chi_2(v,u)}{u^2} - \frac{\alpha\hbar}{2} \pi
\int_{2V_s}^{2 V_\infty} dv \chi_u (v,0)
\nonumber \\
 &=& -\hbar \int_{V_s}^{V_\infty} d \epsilon \int_{V_s}^{V_\infty} d \epsilon'
\frac{F_2^{(ss)}(z,z',\epsilon, \epsilon')}{(\epsilon - \epsilon')^2}
\nonumber \\
& & 
- \alpha \beta  \hbar \pi \int_{V_s}^{V_\infty} d\epsilon g_s^2(\epsilon)
|\psi^{(s)}(\epsilon,z)|^2 |\psi^{(s)}(\epsilon,z')|^2 
f_{FD}( \epsilon- \mu_s).
\end{eqnarray}
Here we omitted the principal value operation in the first integral since
the integrand is regular at $u=0$ because the leading order term in 
$\chi_2(v,u\rightarrow 0)$ is parabolic in $u$.

\bibliography{article}

\begin{thebibliography}{45}
\expandafter\ifx\csname natexlab\endcsname\relax\def\natexlab#1{#1}\fi
\expandafter\ifx\csname bibnamefont\endcsname\relax
  \def\bibnamefont#1{#1}\fi
\expandafter\ifx\csname bibfnamefont\endcsname\relax
  \def\bibfnamefont#1{#1}\fi
\expandafter\ifx\csname citenamefont\endcsname\relax
  \def\citenamefont#1{#1}\fi
\expandafter\ifx\csname url\endcsname\relax
  \def\url#1{\texttt{#1}}\fi
\expandafter\ifx\csname urlprefix\endcsname\relax\def\urlprefix{URL }\fi
\providecommand{\bibinfo}[2]{#2}
\providecommand{\eprint}[2][]{\url{#2}}

\bibitem[{\citenamefont{Platero and Aguado}(2004)}]{platero}
\bibinfo{author}{\bibfnamefont{G.}~\bibnamefont{Platero}} \bibnamefont{and}
  \bibinfo{author}{\bibfnamefont{R.}~\bibnamefont{Aguado}},
  \bibinfo{journal}{Phys. Rep.} \textbf{\bibinfo{volume}{395}},
  \bibinfo{pages}{1} (\bibinfo{year}{2004}).

\bibitem[{\citenamefont{L{\'o}pez et~al.}(2001)\citenamefont{L{\'o}pez, Aguado,
  Platero, and Tejedor}}]{lopez}
\bibinfo{author}{\bibfnamefont{R.}~\bibnamefont{L{\'o}pez}},
  \bibinfo{author}{\bibfnamefont{R.}~\bibnamefont{Aguado}},
  \bibinfo{author}{\bibfnamefont{G.}~\bibnamefont{Platero}}, \bibnamefont{and}
  \bibinfo{author}{\bibfnamefont{C.}~\bibnamefont{Tejedor}},
  \bibinfo{journal}{Phys. Rev. B} \textbf{\bibinfo{volume}{64}},
  \bibinfo{pages}{075319} (\bibinfo{year}{2001}).

\bibitem[{\citenamefont{You et~al.}(2000)\citenamefont{You, Lam, and
  Zheng}}]{you}
\bibinfo{author}{\bibfnamefont{J.~Q.} \bibnamefont{You}},
  \bibinfo{author}{\bibfnamefont{C.-H.} \bibnamefont{Lam}}, \bibnamefont{and}
  \bibinfo{author}{\bibfnamefont{H.~Z.} \bibnamefont{Zheng}},
  \bibinfo{journal}{Phys. Rev. B} \textbf{\bibinfo{volume}{62}},
  \bibinfo{pages}{1978} (\bibinfo{year}{2000}).

\bibitem[{\citenamefont{Zhou et~al.}(2005)\citenamefont{Zhou, Jiang, and
  Cai}}]{Zhou}
\bibinfo{author}{\bibfnamefont{S.}~\bibnamefont{Zhou}},
  \bibinfo{author}{\bibfnamefont{J.~F.} \bibnamefont{Jiang}}, \bibnamefont{and}
  \bibinfo{author}{\bibfnamefont{Q.~Y.} \bibnamefont{Cai}},
  \bibinfo{journal}{Solid-State Electronics} \textbf{\bibinfo{volume}{49}},
  \bibinfo{pages}{1951} (\bibinfo{year}{2005}).

\bibitem[{\citenamefont{Faleev and Stockman}(2002)}]{faleev}
\bibinfo{author}{\bibfnamefont{S.~V.} \bibnamefont{Faleev}} \bibnamefont{and}
  \bibinfo{author}{\bibfnamefont{M.~I.} \bibnamefont{Stockman}},
  \bibinfo{journal}{Phys. Rev. B} \textbf{\bibinfo{volume}{66}},
  \bibinfo{pages}{085318} (\bibinfo{year}{2002}).

\bibitem[{\citenamefont{Aronov et~al.}(1998)\citenamefont{Aronov, Beletskii,
  Berman, Campbell, Doolen, and Dudiy}}]{aronov98}
\bibinfo{author}{\bibfnamefont{I.~E.} \bibnamefont{Aronov}},
  \bibinfo{author}{\bibfnamefont{N.~N.} \bibnamefont{Beletskii}},
  \bibinfo{author}{\bibfnamefont{G.~P.} \bibnamefont{Berman}},
  \bibinfo{author}{\bibfnamefont{D.~K.} \bibnamefont{Campbell}},
  \bibinfo{author}{\bibfnamefont{G.~D.} \bibnamefont{Doolen}},
  \bibnamefont{and} \bibinfo{author}{\bibfnamefont{S.~V.} \bibnamefont{Dudiy}},
  \bibinfo{journal}{Phys. Rev. B} \textbf{\bibinfo{volume}{58}},
  \bibinfo{pages}{9894} (\bibinfo{year}{1998}).

\bibitem[{\citenamefont{Safi and Schulz}(1995)}]{safi}
\bibinfo{author}{\bibfnamefont{I.}~\bibnamefont{Safi}} \bibnamefont{and}
  \bibinfo{author}{\bibfnamefont{H.~J.} \bibnamefont{Schulz}},
  \bibinfo{journal}{Phys. Rev. B} \textbf{\bibinfo{volume}{52}},
  \bibinfo{pages}{R17040} (\bibinfo{year}{1995}).

\bibitem[{\citenamefont{Feiginov}(2001)}]{feiginov01}
\bibinfo{author}{\bibfnamefont{M.~N.} \bibnamefont{Feiginov}},
  \bibinfo{journal}{Appl. Phys. Lett.} \textbf{\bibinfo{volume}{78}},
  \bibinfo{pages}{3301} (\bibinfo{year}{2001}).

\bibitem[{\citenamefont{Landauer}(1957)}]{lanbue1}
\bibinfo{author}{\bibfnamefont{R.}~\bibnamefont{Landauer}},
  \bibinfo{journal}{IBM J. Res. Dev.} \textbf{\bibinfo{volume}{1}},
  \bibinfo{pages}{223} (\bibinfo{year}{1957}).

\bibitem[{\citenamefont{Landauer}(1987)}]{lanbue2}
\bibinfo{author}{\bibfnamefont{R.}~\bibnamefont{Landauer}},
  \bibinfo{journal}{Z. Phys. B} \textbf{\bibinfo{volume}{68}},
  \bibinfo{pages}{217} (\bibinfo{year}{1987}).

\bibitem[{\citenamefont{Fisher and Lee}(1981)}]{lanbue3}
\bibinfo{author}{\bibfnamefont{D.~S.} \bibnamefont{Fisher}} \bibnamefont{and}
  \bibinfo{author}{\bibfnamefont{P.~A.} \bibnamefont{Lee}},
  \bibinfo{journal}{Phys. Rev. B} \textbf{\bibinfo{volume}{23}},
  \bibinfo{pages}{6851} (\bibinfo{year}{1981}).

\bibitem[{\citenamefont{B{\"u}ttiker}(1986)}]{lanbue4}
\bibinfo{author}{\bibfnamefont{M.}~\bibnamefont{B{\"u}ttiker}},
  \bibinfo{journal}{Phys. Rev. Lett.} \textbf{\bibinfo{volume}{57}},
  \bibinfo{pages}{1761} (\bibinfo{year}{1986}).

\bibitem[{\citenamefont{B{\"u}ttiker}(1988)}]{lanbue5}
\bibinfo{author}{\bibfnamefont{M.}~\bibnamefont{B{\"u}ttiker}},
  \bibinfo{journal}{IBM J. Res. Dev.} \textbf{\bibinfo{volume}{32}},
  \bibinfo{pages}{317} (\bibinfo{year}{1988}).

\bibitem[{\citenamefont{Stone and Szafer}(1988)}]{lanbue6}
\bibinfo{author}{\bibfnamefont{A.~D.} \bibnamefont{Stone}} \bibnamefont{and}
  \bibinfo{author}{\bibfnamefont{A.}~\bibnamefont{Szafer}},
  \bibinfo{journal}{IBM J. Res. Dev.} \textbf{\bibinfo{volume}{32}},
  \bibinfo{pages}{384} (\bibinfo{year}{1988}).

\bibitem[{\citenamefont{B{\"u}ttiker
  et~al.}(1993{\natexlab{a}})\citenamefont{B{\"u}ttiker, Pr{\^{e}}tre, and
  Thomas}}]{buettiker93a}
\bibinfo{author}{\bibfnamefont{M.}~\bibnamefont{B{\"u}ttiker}},
  \bibinfo{author}{\bibfnamefont{A.}~\bibnamefont{Pr{\^{e}}tre}},
  \bibnamefont{and} \bibinfo{author}{\bibfnamefont{H.}~\bibnamefont{Thomas}},
  \bibinfo{journal}{Phys. Rev. Lett.} \textbf{\bibinfo{volume}{70}},
  \bibinfo{pages}{4114} (\bibinfo{year}{1993}{\natexlab{a}}).

\bibitem[{\citenamefont{B{\"u}ttiker}(1993)}]{buettiker93b}
\bibinfo{author}{\bibfnamefont{M.}~\bibnamefont{B{\"u}ttiker}},
  \bibinfo{journal}{J. Phys. C} \textbf{\bibinfo{volume}{5}},
  \bibinfo{pages}{9361} (\bibinfo{year}{1993}).

\bibitem[{\citenamefont{B{\"u}ttiker
  et~al.}(1993{\natexlab{b}})\citenamefont{B{\"u}ttiker, Thomas, and
  Pr{\^{e}}tre}}]{buettiker93c}
\bibinfo{author}{\bibfnamefont{M.}~\bibnamefont{B{\"u}ttiker}},
  \bibinfo{author}{\bibfnamefont{H.}~\bibnamefont{Thomas}}, \bibnamefont{and}
  \bibinfo{author}{\bibfnamefont{A.}~\bibnamefont{Pr{\^{e}}tre}},
  \bibinfo{journal}{Physics Letters A} \textbf{\bibinfo{volume}{180}},
  \bibinfo{pages}{364} (\bibinfo{year}{1993}{\natexlab{b}}).

\bibitem[{\citenamefont{B{\"u}ttiker et~al.}(1994)\citenamefont{B{\"u}ttiker,
  Thomas, and Pr{\^{e}}tre}}]{buettiker94a}
\bibinfo{author}{\bibfnamefont{M.}~\bibnamefont{B{\"u}ttiker}},
  \bibinfo{author}{\bibfnamefont{H.}~\bibnamefont{Thomas}}, \bibnamefont{and}
  \bibinfo{author}{\bibfnamefont{A.}~\bibnamefont{Pr{\^{e}}tre}},
  \bibinfo{journal}{Z. Phys. B} \textbf{\bibinfo{volume}{94}},
  \bibinfo{pages}{133} (\bibinfo{year}{1994}).

\bibitem[{\citenamefont{B{\"u}ttiker and Christen}(1996)}]{buettiker96a}
\bibinfo{author}{\bibfnamefont{M.}~\bibnamefont{B{\"u}ttiker}}
  \bibnamefont{and} \bibinfo{author}{\bibfnamefont{T.}~\bibnamefont{Christen}},
  in \emph{\bibinfo{booktitle}{Quantum Transport in Semiconductor Submicron
  Structures}}, edited by
  \bibinfo{editor}{\bibfnamefont{B.}~\bibnamefont{Kramer}}
  (\bibinfo{publisher}{Kluwer Academic Publishers},
  \bibinfo{address}{Dodrecht}, \bibinfo{year}{1996}), p. \bibinfo{pages}{263}.

\bibitem[{\citenamefont{B{\"u}ttiker}(1996)}]{buettiker96b}
\bibinfo{author}{\bibfnamefont{M.}~\bibnamefont{B{\"u}ttiker}},
  \bibinfo{journal}{J. Math. Phys.} \textbf{\bibinfo{volume}{37}},
  \bibinfo{pages}{4793} (\bibinfo{year}{1996}).

\bibitem[{\citenamefont{Christen and
  B{\"u}ttiker}(1996{\natexlab{a}})}]{christen96a}
\bibinfo{author}{\bibfnamefont{T.}~\bibnamefont{Christen}} \bibnamefont{and}
  \bibinfo{author}{\bibfnamefont{M.}~\bibnamefont{B{\"u}ttiker}},
  \bibinfo{journal}{Phys. Rev. B} \textbf{\bibinfo{volume}{53}},
  \bibinfo{pages}{2064} (\bibinfo{year}{1996}{\natexlab{a}}).

\bibitem[{\citenamefont{Christen and
  B{\"u}ttiker}(1996{\natexlab{b}})}]{christen96b}
\bibinfo{author}{\bibfnamefont{T.}~\bibnamefont{Christen}} \bibnamefont{and}
  \bibinfo{author}{\bibfnamefont{M.}~\bibnamefont{B{\"u}ttiker}},
  \bibinfo{journal}{Phys. Rev. Lett.} \textbf{\bibinfo{volume}{77}},
  \bibinfo{pages}{143} (\bibinfo{year}{1996}{\natexlab{b}}).

\bibitem[{\citenamefont{Pr{\^{e}}tre
  et~al.}(1996{\natexlab{a}})\citenamefont{Pr{\^{e}}tre, Thomas, and
  B{\"u}ttiker}}]{pretre96}
\bibinfo{author}{\bibfnamefont{A.}~\bibnamefont{Pr{\^{e}}tre}},
  \bibinfo{author}{\bibfnamefont{H.}~\bibnamefont{Thomas}}, \bibnamefont{and}
  \bibinfo{author}{\bibfnamefont{M.}~\bibnamefont{B{\"u}ttiker}},
  \bibinfo{journal}{Phys. Rev. B} \textbf{\bibinfo{volume}{54}},
  \bibinfo{pages}{8130} (\bibinfo{year}{1996}{\natexlab{a}}).

\bibitem[{\citenamefont{Gopar et~al.}(1996)\citenamefont{Gopar, Mello, and
  B{\"u}ttiker}}]{gopar96}
\bibinfo{author}{\bibfnamefont{V.~A.} \bibnamefont{Gopar}},
  \bibinfo{author}{\bibfnamefont{P.~A.} \bibnamefont{Mello}}, \bibnamefont{and}
  \bibinfo{author}{\bibfnamefont{M.}~\bibnamefont{B{\"u}ttiker}},
  \bibinfo{journal}{Phys. Rev. Lett.} \textbf{\bibinfo{volume}{77}},
  \bibinfo{pages}{3005} (\bibinfo{year}{1996}).

\bibitem[{\citenamefont{Brouwer and B{\"u}ttiker}(1997)}]{brouwer97}
\bibinfo{author}{\bibfnamefont{P.~W.} \bibnamefont{Brouwer}} \bibnamefont{and}
  \bibinfo{author}{\bibfnamefont{M.}~\bibnamefont{B{\"u}ttiker}},
  \bibinfo{journal}{Europhys. Lett.} \textbf{\bibinfo{volume}{37}},
  \bibinfo{pages}{441} (\bibinfo{year}{1997}).

\bibitem[{\citenamefont{Blanter and B{\"u}ttiker}(2000)}]{blanter_pr00}
\bibinfo{author}{\bibfnamefont{Y.~M.} \bibnamefont{Blanter}} \bibnamefont{and}
  \bibinfo{author}{\bibfnamefont{M.}~\bibnamefont{B{\"u}ttiker}},
  \bibinfo{journal}{Phys. Rep.} \textbf{\bibinfo{volume}{336}},
  \bibinfo{pages}{1} (\bibinfo{year}{2000}).

\bibitem[{\citenamefont{Hekking and Pekola}(2006)}]{hekking}
\bibinfo{author}{\bibfnamefont{F.~W.~J.} \bibnamefont{Hekking}}
  \bibnamefont{and} \bibinfo{author}{\bibfnamefont{J.~P.}
  \bibnamefont{Pekola}}, \bibinfo{journal}{Phys. Rev. Lett.}
  \textbf{\bibinfo{volume}{96}}, \bibinfo{pages}{056603}
  (\bibinfo{year}{2006}).

\bibitem[{\citenamefont{Fu and Dudley}(1993)}]{fu}
\bibinfo{author}{\bibfnamefont{Y.}~\bibnamefont{Fu}} \bibnamefont{and}
  \bibinfo{author}{\bibfnamefont{S.~C.} \bibnamefont{Dudley}},
  \bibinfo{journal}{Phys. Rev. Lett.} \textbf{\bibinfo{volume}{70}},
  \bibinfo{pages}{65} (\bibinfo{year}{1993}).

\bibitem[{\citenamefont{Blanter and B{\"u}ttiker}(1998)}]{blanter98a}
\bibinfo{author}{\bibfnamefont{Y.~M.} \bibnamefont{Blanter}} \bibnamefont{and}
  \bibinfo{author}{\bibfnamefont{M.}~\bibnamefont{B{\"u}ttiker}},
  \bibinfo{journal}{Europhys. Lett.} \textbf{\bibinfo{volume}{42}},
  \bibinfo{pages}{535} (\bibinfo{year}{1998}).

\bibitem[{\citenamefont{Blanter et~al.}(1998)\citenamefont{Blanter, Hekking,
  and B{\"u}ttiker}}]{blanter98b}
\bibinfo{author}{\bibfnamefont{Y.~M.} \bibnamefont{Blanter}},
  \bibinfo{author}{\bibfnamefont{F.~W.~J.} \bibnamefont{Hekking}},
  \bibnamefont{and}
  \bibinfo{author}{\bibfnamefont{M.}~\bibnamefont{B{\"u}ttiker}},
  \bibinfo{journal}{Phys. Rev. Lett.} \textbf{\bibinfo{volume}{81}},
  \bibinfo{pages}{1925} (\bibinfo{year}{1998}).

\bibitem[{\citenamefont{Wulf et~al.}(2003)\citenamefont{Wulf, Racec, Racec, and
  Aldea}}]{emrs03}
\bibinfo{author}{\bibfnamefont{U.}~\bibnamefont{Wulf}},
  \bibinfo{author}{\bibfnamefont{E.~R.} \bibnamefont{Racec}},
  \bibinfo{author}{\bibfnamefont{P.~N.} \bibnamefont{Racec}}, \bibnamefont{and}
  \bibinfo{author}{\bibfnamefont{A.}~\bibnamefont{Aldea}},
  \bibinfo{journal}{Mat. Sci. Eng. C} \textbf{\bibinfo{volume}{23}},
  \bibinfo{pages}{675} (\bibinfo{year}{2003}).

\bibitem[{\citenamefont{Racec and Wulf}(2006)}]{emrs05}
\bibinfo{author}{\bibfnamefont{P.~N.} \bibnamefont{Racec}} \bibnamefont{and}
  \bibinfo{author}{\bibfnamefont{U.}~\bibnamefont{Wulf}},
  \bibinfo{journal}{Mat. Sci. Eng. C} \textbf{\bibinfo{volume}{26}},
  \bibinfo{pages}{876} (\bibinfo{year}{2006}).

\bibitem[{\citenamefont{Baym and Kadanoff}(1961)}]{kadanoff}
\bibinfo{author}{\bibfnamefont{G.}~\bibnamefont{Baym}} \bibnamefont{and}
  \bibinfo{author}{\bibfnamefont{L.~P.} \bibnamefont{Kadanoff}},
  \bibinfo{journal}{Phys. Rev.} \textbf{\bibinfo{volume}{124}},
  \bibinfo{pages}{287} (\bibinfo{year}{1961}).

\bibitem[{\citenamefont{Dolgopolov et~al.}(1996)\citenamefont{Dolgopolov,
  Shashkin, Aristov, Scherek, Drexler, Hansen, Kotthaus, and
  Holland}}]{dolgopolov}
\bibinfo{author}{\bibfnamefont{V.~T.} \bibnamefont{Dolgopolov}},
  \bibinfo{author}{\bibfnamefont{A.~A.} \bibnamefont{Shashkin}},
  \bibinfo{author}{\bibfnamefont{A.~V.} \bibnamefont{Aristov}},
  \bibinfo{author}{\bibfnamefont{D.}~\bibnamefont{Scherek}},
  \bibinfo{author}{\bibfnamefont{H.}~\bibnamefont{Drexler}},
  \bibinfo{author}{\bibfnamefont{W.}~\bibnamefont{Hansen}},
  \bibinfo{author}{\bibfnamefont{J.~P.} \bibnamefont{Kotthaus}},
  \bibnamefont{and} \bibinfo{author}{\bibfnamefont{M.}~\bibnamefont{Holland}},
  \bibinfo{journal}{Phys. Low-Dim. Struct.} \textbf{\bibinfo{volume}{6}},
  \bibinfo{pages}{1} (\bibinfo{year}{1996}).

\bibitem[{\citenamefont{Racec et~al.}(2003)\citenamefont{Racec, Racec, and
  Wulf}}]{roxana}
\bibinfo{author}{\bibfnamefont{E.~R.} \bibnamefont{Racec}},
  \bibinfo{author}{\bibfnamefont{P.~N.} \bibnamefont{Racec}}, \bibnamefont{and}
  \bibinfo{author}{\bibfnamefont{U.}~\bibnamefont{Wulf}},
  \bibinfo{journal}{Recent Res. Devel. Physics} \textbf{\bibinfo{volume}{4}},
  \bibinfo{pages}{387} (\bibinfo{year}{2003}), \eprint{cond-mat/0404425}.

\bibitem[{\citenamefont{Ehrenreich and Cohen}(1959)}]{ehrenreich}
\bibinfo{author}{\bibfnamefont{H.}~\bibnamefont{Ehrenreich}} \bibnamefont{and}
  \bibinfo{author}{\bibfnamefont{M.~H.} \bibnamefont{Cohen}},
  \bibinfo{journal}{Phys. Rev.} \textbf{\bibinfo{volume}{115}},
  \bibinfo{pages}{786} (\bibinfo{year}{1959}).

\bibitem[{\citenamefont{Haken}(1976)}]{haken}
\bibinfo{author}{\bibfnamefont{H.}~\bibnamefont{Haken}},
  \emph{\bibinfo{title}{Quantum Field Theory of Solids}}
  (\bibinfo{publisher}{Elsevier Science Publishing Company},
  \bibinfo{address}{North-Holland}, \bibinfo{year}{1976}).

\bibitem[{\citenamefont{Hedin and Lundquist}(1969)}]{hedin}
\bibinfo{author}{\bibfnamefont{L.}~\bibnamefont{Hedin}} \bibnamefont{and}
  \bibinfo{author}{\bibfnamefont{S.}~\bibnamefont{Lundquist}}, in
  \emph{\bibinfo{booktitle}{Solid State Physics}} (\bibinfo{publisher}{Academic
  Press}, \bibinfo{address}{New York}, \bibinfo{year}{1969}),
  vol.~\bibinfo{volume}{23}.

\bibitem[{\citenamefont{Racec et~al.}(2000)\citenamefont{Racec, Wulf, and
  Ku{\v{c}}era}}]{sse00}
\bibinfo{author}{\bibfnamefont{P.~N.} \bibnamefont{Racec}},
  \bibinfo{author}{\bibfnamefont{U.}~\bibnamefont{Wulf}}, \bibnamefont{and}
  \bibinfo{author}{\bibfnamefont{J.}~\bibnamefont{Ku{\v{c}}era}},
  \bibinfo{journal}{Solid-State Electron.} \textbf{\bibinfo{volume}{44}},
  \bibinfo{pages}{881} (\bibinfo{year}{2000}).

\bibitem[{\citenamefont{Racec et~al.}(2002)\citenamefont{Racec, Racec, and
  Wulf}}]{prb02}
\bibinfo{author}{\bibfnamefont{P.~N.} \bibnamefont{Racec}},
  \bibinfo{author}{\bibfnamefont{E.~R.} \bibnamefont{Racec}}, \bibnamefont{and}
  \bibinfo{author}{\bibfnamefont{U.}~\bibnamefont{Wulf}},
  \bibinfo{journal}{Phys. Rev. B} \textbf{\bibinfo{volume}{65}},
  \bibinfo{pages}{193314} (\bibinfo{year}{2002}).

\bibitem[{\citenamefont{Pr{\^{e}}tre
  et~al.}(1996{\natexlab{b}})\citenamefont{Pr{\^{e}}tre, Thomas, and
  B{\"u}ttiker}}]{buettiker96c}
\bibinfo{author}{\bibfnamefont{A.}~\bibnamefont{Pr{\^{e}}tre}},
  \bibinfo{author}{\bibfnamefont{H.}~\bibnamefont{Thomas}}, \bibnamefont{and}
  \bibinfo{author}{\bibfnamefont{M.}~\bibnamefont{B{\"u}ttiker}},
  \bibinfo{journal}{Phys. Rev. B} \textbf{\bibinfo{volume}{54}},
  \bibinfo{pages}{8130} (\bibinfo{year}{1996}{\natexlab{b}}).

\bibitem[{\citenamefont{B{\"u}ttiker}(1999)}]{buttiker98}
\bibinfo{author}{\bibfnamefont{M.}~\bibnamefont{B{\"u}ttiker}},
  \bibinfo{journal}{J. Korean Phys. Soc.} \textbf{\bibinfo{volume}{34}},
  \bibinfo{pages}{S121} (\bibinfo{year}{1999}).

\bibitem[{\citenamefont{Nolting}(1992)}]{nolting7}
\bibinfo{author}{\bibfnamefont{W.}~\bibnamefont{Nolting}},
  \emph{\bibinfo{title}{Grundkurs: Theoretische Physik 7}}
  (\bibinfo{publisher}{Verlag Zimmermann-Neufang}, \bibinfo{address}{Ulmen},
  \bibinfo{year}{1992}), chap. \bibinfo{chapter}{1.2}.

\bibitem[{\citenamefont{Czycholl}(2000)}]{czy}
\bibinfo{author}{\bibfnamefont{G.}~\bibnamefont{Czycholl}},
  \emph{\bibinfo{title}{Theoretische Festk{\"o}rperphysik}}
  (\bibinfo{publisher}{Fried. Vieweg \& Sohn},
  \bibinfo{address}{Braunschweig/Wiesbaden}, \bibinfo{year}{2000}), chap.
  \bibinfo{chapter}{7.6}.

\bibitem[{\citenamefont{Gelfand and Schilow}(1960)}]{gelfand}
\bibinfo{author}{\bibfnamefont{I.~M.} \bibnamefont{Gelfand}} \bibnamefont{and}
  \bibinfo{author}{\bibfnamefont{G.~E.} \bibnamefont{Schilow}},
  \emph{\bibinfo{title}{Verallgemeinerte Funktionen (Distributionen) I}}
  (\bibinfo{publisher}{Deutscher Verlag der Wissenschaften},
  \bibinfo{address}{Berlin}, \bibinfo{year}{1960}).

\end{thebibliography}

\end{document}